\documentclass[11pt]{iopart}
\usepackage{iopams,graphicx,cite,color,nicefrac,url}

\newcommand{\eps}{\epsilon}
\newcommand{\REF}[1]{\textcolor{red}{REF}}
\newcommand{\avg}[1]{\left\langle #1 \right\rangle}

\begin{document}

\title[~]{Optimal metabolic strategies for microbial growth in stationary random environments}

\author{Anna Paola Muntoni and Andrea De Martino}

\address{Politecnico di Torino, Turin, Italy, and Italian Institute for Genomic Medicine, Turin, Italy}
\ead{andrea.demartino@polito.it}
\vspace{10pt}

\begin{abstract}
In order to grow in any given environment, bacteria need to collect information about the medium composition and implement suitable growth strategies by adjusting their regulatory and metabolic degrees of freedom. In the standard sense, optimal strategy selection is achieved when bacteria grow at the fastest rate possible in that medium. While this view of optimality is well suited for cells that have perfect knowledge about their surroundings (e.g. nutrient levels), things are more involved in uncertain or fluctuating conditions, especially when changes occur over timescales comparable to (or faster than) those required to organize a response. Information theory however provides recipes for how cells can choose the optimal growth strategy under uncertainty about the stress levels they will face. Here we analyse the theoretically optimal scenarios for a coarse-grained, experiment-inspired model of bacterial metabolism for growth in a medium described by the (static) probability density of a single variable (the `stress level'). We show that heterogeneity in growth rates consistently emerges as the optimal response when the environment is sufficiently complex and/or when perfect adjustment of metabolic degrees of freedom is not possible (e.g. due to limited resources). In addition, outcomes close to those achievable with unlimited resources are often attained effectively with a modest amount of fine tuning. In other terms, heterogeneous population structures in complex media may be rather robust with respect to the resources available to probe the environment and adjust reaction rates. 
\end{abstract}

%
%
%
%
%

\section{Introduction}


The standard theoretical view of bacterial growth posits that, in any growth medium, cells are capable of adjusting their metabolic degrees of freedom (i.e. the rates of metabolic reactions) within bounds dictated by thermodynamic and regulatory constraints (e.g. enzyme expression levels, reaction free energies, etc.) so as to maximize their growth rate \cite{searching}. Besides evolutionary considerations, such a picture is supported by the fact that the expression levels of certain metabolic enzymes and of basic macromolecular machines like ribosomes actually appear to be tuned for growth-rate maximization in bacterial populations \cite{control,howfast}. On the other hand, the significant cell-to-cell variability in growth rates observed in experiments \cite{evidencefora,stochasticity,cellsize,individuality,fundamental}, together with the fact that constraints arising outside metabolism suffice to explain a large batch of empirical facts without assuming any growth-rate optimization \cite{interdependence,threshold}, appears to call for deeper insight into the notion of `optimality' for bacterial growth. 

Recent work has shown that the relationship between population growth and cell-to-cell variability is well described by a Maximum Entropy (MaxEnt) theory leading to a variable trade-off akin to the usual energy-entropy balance in statistical physics. More specifically, {\it E. coli} populations growing in carbon-limited media realize a close-to-optimal fitness-heterogeneity trade-off in rich media \cite{growthvs}, while they seem to be less variable or slower-growing than optimal in poorer growth conditions \cite{relationship}. Metabolic fluxes likewise appear to be better captured by accounting for such a trade-off than by a standard optimality assumption \cite{statmech}. In each case, the balance between growth and variability is described by a finite (medium-dependent) `temperature', where a zero-temperature (resp. infinite-temperature) limit corresponds to maximal growth (resp. maximal variability). Save for a few general ideas derived from broad-brush models \cite{quantifying}, what determines the `temperature' (i.e. the  fitness-heterogeneity balance) of actual microbial systems is still unclear. 

High variability can naturally arise from unavoidable inter-cellular differences in gene expression levels or regulatory programs (e.g. cell cycle) \cite{noisein}. In models of metabolism, this would lead, at the simplest level, to cell-dependent changes in the constraints under which growth is optimized. In this respect, cell-to-cell heterogeneity might be interpreted as `optimality plus noise', and the `temperature' described above would quantify, in essence, the noise strength. Importantly, though, there might be an inherent advantage in maintaining a diverse population, especially in environments that fluctuate (due e.g. to natural variability) or when cells have imperfect information about their growth medium (due e.g. to limits in precision caused by the high costs cells face to maintain and operate a sensing apparatus). These factors are not usually included in standard models of metabolic networks, which therefore cannot  address the fitness benefits of heterogeneity.

The problem of growth maximization clearly becomes more subtle under uncertainty about  environmental conditions, as the straightforward optimization that can be carried out in a perfectly known medium is no longer an option. Recipes for selecting the optimal growth strategy in uncertain environments are, however, provided by information theory. Theoretical work aimed at understanding how efficiently populations can harvest, process and exploit information about variable or unpredictable media has indeed shown that {\it bet-hedging} (i.e. maintaining a fraction of slow-growing cells even in rich media or sustaining a lower short-term growth to ensure faster long-term growth) can yield significant fitness gains in a wide variety of situations \cite{thevalue}. Biological implications of these results have been explored against several backdrops \cite{exex,nongenetic,pareto}, albeit never specifically in the context of metabolism. 

In this paper, inspired by the above studies as well as by \cite{yieldcost} and \cite{textbook} (Chapter 6), we look at a minimal, experiment-derived mathematical model to characterize optimal metabolic strategies for growth in uncertain environments, focusing for simplicity on static environments defined by the probability density of a single variable (the `stress level'). Optimal strategies are parametrised by a `temperature' that modulates the amount of information they encode about the growth medium or, loosely speaking, how precisely cells can match their phenotype to the external conditions in order to foster growth. We will show explicitly that, when the medium is sufficiently complex, optimal populations acquire a non-trivial phenotypic organization even when the metabolic strategy encodes the maximum possible amount of information about the external conditions. A broad spectrum of behaviours is uncovered upon varying the structure of the environment. Remarkably, however, the emerging scenarios are often robust to changes in the `temperature'. This suggests that metabolic networks may yield outcomes close to globally optimal ones (at least in an information-theoretic sense) even when resources to probe the environment and adjust metabolic reactions are limited.

\section{Results}

\subsection{Model of metabolism and growth}

We consider a coarse-grained model of microbial growth metabolism in which each cell's metabolic strategy is described by just two quantities, namely the specific uptake (or inverse growth yield) $q$, quantifying the nutrient intake required to grow per unit of growth rate, and the biosynthetic expenditure $\eps$, quantifying the proteome mass fraction to be devoted to metabolic enzymes per unit of growth rate. (In more detailed models of metabolic networks, the former quantity relates to the rate at which the limiting nutrient is imported, while the latter is proportional to a weighted sum of the absolute values of the fluxes through metabolic reactions \cite{cafba,yieldcost}.) For {\it E. coli} growing in carbon-limited media it has been argued that, for given $q$ and $\epsilon$, the growth rate $\mu$ is well described by the formula \cite{yieldcost}
\begin{eqnarray}\label{groo}
\mu \simeq\frac{\phi}{w+s q+\eps}~~,
\end{eqnarray}
where $s\geq 0$ represents the level of nutritional stress to which the organism is subject, while $\phi>0$ and $w>0$ are constants representing respectively the fraction of proteome devoted to constitutively expressed proteins and the proteome share to be allocated to ribosome-affiliated proteins per unit of growth rate. (For glucose-limited {\it E. coli} growth, $\phi\simeq 0.48$ and $w\simeq 0.169$ h \cite{interdependence}.) For our purposes, $s$ can be assumed to be inversely proportional to the carbon level as argued in \cite{cafba} (so $s\ll 1$ and $s\gg 1$ for carbon-rich and carbon-poor environments, respectively).

\begin{figure}
\begin{center}
\includegraphics[width=\linewidth]{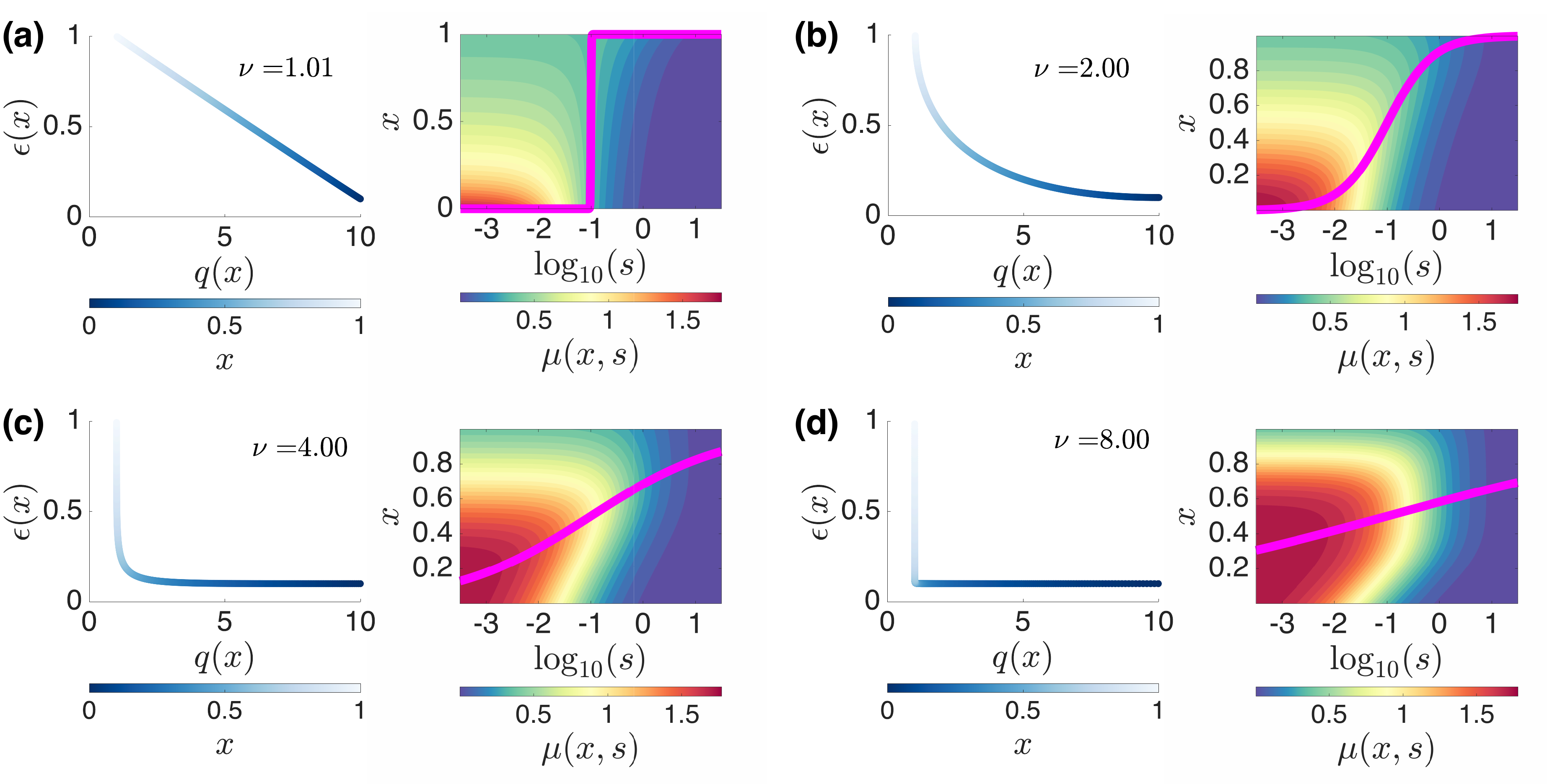}
\caption{Representative scenarios for $q\left(x\right)$ vs $\eps\left(x\right)$ (Equations (\ref{eq:q_q}) and (\ref{eq:q_eps}), left plots in each panel) with the corresponding fitness landscape $\mu(x,s)$ (right plots), for $\nu=1.01$ (panel (a)), 2 (b), 4 (c) and 8 (d). The values of $x$ that maximize $\mu$ for each $s$, i.e. $\widehat{x}(s)$ (Equation (\ref{eq:x_hat})), are represented by a magenta line. The constants characterizing the $R$ and $F$ strategies are set to $q_F=10$, $q_R=1$, $\eps_F=0.1$ and $\eps_R=1$. \label{fig:fig1}}
\end{center}
\end{figure}

Let us assume that the stress level $s$ in (\ref{groo}) is a homogeneous parameter whose value is controlled externally. If $\mu$ were to be maximized, the quantity $sq+\eps$ would have to be minimized. In carbon-limited {\it E. coli} growth, however, $q$ and $\epsilon$ are subject to a trade-off such that high $q$ implies low $\eps$ and vice versa \cite{yieldcost}. Metabolic states with minimal biosynthetic expenditure are hence favoured in rich environments (small $s$), while states of minimal nutritional requirements prevail in poor media (large $s$). To make a concrete model, we draw inspiration from the fermentation/respiration duality that again characterizes {\it E. coli} growth in carbon-limited media and assume that the organism can regulate $q$ and $\epsilon$ between two extreme strategies, denoted by indices $F$ and $R$, respectively, distinguished by the fact that $q_F>q_R$ (i.e. the specific nutrient requirements of $F$ are higher than those of $R$) and $\eps_R>\eps_F$ (i.e. the specific expenditure for $R$ is higher than for $F$). (Estimated values of the specific nutrient intake and the specific proteome mass fraction devoted to metabolic enzymes required for {\it E.coli} to grow on lactate under fermentation are $q_F\simeq 8$\,g$_{\mathrm{lac}}/$g$_{\mathrm{DW}}$ and $\eps_F\simeq 0.3$\,h, respectively; the corresponding quantities under respiration are instead given by $q_R\simeq 5$\,g$_{\mathrm{lac}}/$g$_{\mathrm{DW}}$ and $\eps_R\simeq 0.6$\,h \cite{yieldcost}. In this work, we choose the representative values $q_F= 10$, $\eps_F= 0.1$, $q_R= 1$, $\eps_R= 1$, omitting the units as the specifics of the carbon source are immaterial for us. Nevertheless, with these choices the growth rate $\mu$ can be interpreted to be measured in $1/$h.) We then describe the trade-off between $q$ and $\epsilon$ by assuming that both depend on a single variable $x$ whose value ranges between 0 and 1, such that 
\begin{eqnarray}\label{eq:q_q}
q(x)=q_R+(q_F-q_R)(1-x)^\nu~~,\\
\eps(x)=\eps_F+(\eps_R-\eps_F)x^\nu~~, \label{eq:q_eps}
\end{eqnarray}
where $\nu>1$ is a constant. The growth rate of the organism will then be given by 
\begin{eqnarray}
\mu(x,s)=\frac{\phi}{w+s q(x)+\eps(x)}~~. \label{eq:fitness}
\end{eqnarray}
By taking the derivative of $\mu(x,s)$ over $x$ at fixed $s$, one finds that, for any given $q(x)$ and $\eps(x)$, $\mu$ is maximum when
\begin{eqnarray}
s\frac{\partial q}{\partial x}+\frac{\partial \eps}{\partial x}=0~~.
\end{eqnarray}
If $q(x)$ and $\eps(x)$ are given by (\ref{eq:q_q}) and (\ref{eq:q_eps}), the maximum is achieved for $x=\widehat{x}(s)$, with
\begin{eqnarray}
\widehat{x}(s)=\frac{s^n}{s^n+s_c^n}
~~~~~,~~~~~ s_c=\frac{\eps_R-\eps_F}{q_F-q_R}~~~,~~~ n=\frac{1}{\nu-1}~~. \label{eq:x_hat}
\end{eqnarray}
This means that media with $s\ll s_c$ can be considered to be rich, while media with $s\gg s_c$ are effectively poor. (With our choices for the parameters, $s_c=0.1$.) In turn, if $\mu$ is maximized, strategy $F$ (i.e. $x=0$) will be used in rich environments (where $\eps$ should be as small as possible) while strategy $R$ (i.e. $x=1$) will be used in poor ones (where $q$ should be as small as possible), as shown in Figure \ref{fig:fig1}.
As one modulates the stress level between these two extremes, growth is maximized by intermediate values of $x$ (i.e. by strategies that use both $F$ and $R$). Other choices generically lead to slower growth. 

Notice that the $q-\eps$ trade-off gets stronger as $\nu$ approaches $1$, when the corresponding optimal strategy is a step-like function. Conversely, it gets weaker and weaker as $\nu$ increases. For sakes of simplicity, in the following we shall always use the value $\nu=3/2$, which qualitatively reproduces the trade-off reconstructed from empirical data \cite{yieldcost}. A discussion of how results depend on $\nu$ (including the issue of why a specific value of $\nu$ may be evolutionarily preferred) is deferred to future work.

\subsection{Optimizing growth in random environments: theoretical framework}

Consider a microbe whose growth rate depends on $x$ and $s$ as in  (\ref{eq:fitness}). For any fixed $s$, the cell can maximize $\mu$ by setting $x=\widehat{x}(s)$ (see (\ref{eq:x_hat})). Suppose, however, that $s$ is a random variable with a prescribed probability density $p(s)$. In this case, cells face an uncertainty about the exact value of $s$ they will encounter, although they have knowledge of the {\it ensemble} of environmental conditions in which they live (i.e., of $p(s)$). What is the optimal choice for $x$ in this situation? The most sensible measure of performance is now arguably given by the mean growth rate 
\begin{eqnarray}
\avg{\mu}=\int ds\, p(s)\int dx\, p(x|s) \mu(x,s)~~,
\end{eqnarray}
whose value depends on the conditional distribution $p(x|s)$ that describes the stochastic rule used by the cell to select $x$ for any $s$. The question of optimality therefore concerns the optimal choice of $p(x|s)$. It turns out that this choice depends on the amount of information about $s$ that is encoded in $x$. In particular, if 
\begin{eqnarray}
I=\int ds\, p(s)\int dx\, p(x|s)\log_2\frac{p(x|s)}{p(x)}~~, \label{eq:info}
\end{eqnarray}
denotes the mutual information of $x$ and $s$ (in bits), and 
\begin{eqnarray}
p(x)=\int ds\, p(s) p(x|s)~~,\label{qa}
\end{eqnarray}
then the optimal $p(x|s)$ is given by the solution of 
\begin{eqnarray}
\max_{p(x|s)}\avg{\mu}~~~\textrm{subject to}~~~ 
I=\mathrm{constant}~~, 
\end{eqnarray}
i.e. (see Appendix and \cite{textbook})
\begin{eqnarray}\label{pags}
p^\star(x|s)=\frac{p^{\star}(x)}{N(s,\beta)}\,\, e^{\beta \mu(x,s)}~~,
\end{eqnarray}
where
\begin{eqnarray}
N(s,\beta)=\int dx\, p^{\star}(x) \,e^{\beta \mu(x,s)} \label{pags2}~~,
\end{eqnarray}
while the `inverse temperature' $\beta$ is a Lagrange multiplier and $p^\star(x)$ denotes the probability density of $x$ corresponding to the optimal choice. Eq. (\ref{pags}) has a rather straightforward interpretation. When $\beta\to 0$ (`infinite temperature'), the choice of $x$ becomes independent of $s$, which implies $I=0$. As $\beta$ increases, $I$ increases (i.e. cellular responses encode more and more information about the environment) and $p^\star(x|s)$ tends to get more and more sharply peaked around $\widehat{x}(s)$. For $\beta\to\infty$ (`zero temperature'), in particular, one gets $p^\star(x|s)\simeq\delta[x-\widehat{x}(s)]$, so that cells respond to each instance of $s$ by exactly choosing the value of $x$ that maximizes $\mu$. Such an `infinite-precision' response requires $I$ to be maximal. In other terms, the higher $I$, the higher $\avg{\mu}$. (Conversely, as $\beta$ becomes more and more negative, $p^\star(x|s)$ tends to get more and more sharply peaked around the value of $x$ that {\it minimizes} growth for any $s$. For sake of clarity, we shall however limit the following analysis to the case $\beta> 0$.) $\beta$ is therefore a parameter by which one interpolates between the case of optimal response to any environmental cue and that in which the cell's metabolic strategy is completely insensitive to $s$. Note that, at optimality, the mean growth rate and mutual information given by
\begin{eqnarray}
\langle \mu \rangle_\star&=&\int ds\, p(s)\int dx\, p^\star(x|s) \mu(x,s)~~,\\
I^\star&=&\int ds\, p(s)\int dx\, p^\star(x|s)\log_2\frac{p^\star(x|s)}{p^\star(x)}~~,
\end{eqnarray}
are both functions of $\beta$ and are related by
\begin{equation}
I^{\star} = \frac{\beta}{\ln 2} \langle \mu\rangle_\star - \langle \log_{2} N(s,\beta) \rangle_\star~~.\label{eq:gr_I} 
\end{equation}

For different choices of $p(s)$ and  $\beta$ one can solve Equations (\ref{qa}), (\ref{pags}) and (\ref{pags2}) numerically by iteration from an initial guess. Starting (iteration $n=0$) from uniform guesses for $p\left(x\right)$ and $p\left(x | s\right)$, we iterated Equations (\ref{pags2}), (\ref{qa}), and (\ref{pags}) up to numerical convergence, which we assumed to be achieved  when 
\begin{eqnarray}
&\max_{x} | \,p^{(n)}(x) - p^{(n-1)}(x) |<\sigma ~~~\mathrm{and}~~\\
&\max_{x} | \,p^{(n)}(x | s) - p^{(n-1)}(x | s) |<\sigma~~, 
\end{eqnarray}
where the index $n$ denotes the iteration step while $\sigma$ a numerical precision threshold. (A Matlab code implementing the above procedure is available from \url{https://github.com/anna-pa-m/OptMetStrategy}.) Solutions will clarify how optimal metabolic strategy and population structure change with $\beta$, i.e. with the amount of information about the environment encoded in $x$, in any given environmental condition (described by the chosen $p(s)$). 

Before moving on, we note that the above setting suggests how an `optimal' value of $\beta$ may arise in this scenario. It is indeed reasonable to think that $I^\star$ is directly related to the quantity of cellular resources devoted to probing the environment and tuning metabolic reactions, and, in turn, that higher costs for sensing and metabolism may negatively affect the fitness if resources are limited. If one assumes for simplicity that such a cost reduces the growth performance by a constant amount $c$ per bit of information encoded, the fitness faced by the organism can be written as 
\begin{equation}\label{resfit}
{\cal F}=\avg{\mu}_\star-cI^\star~~. 
\end{equation}
One easily sees that, contrary to $\avg{\mu}_\star$ and $I^\star$ (both of which increase steadily with $\beta$), ${\cal F}$ is maximum when $c=\partial\avg{\mu}_\star/\partial I^\star$, i.e. for 
\begin{equation}\label{betasta}
\beta=\beta^\star\equiv\frac{\ln 2}{c}~~.
\end{equation}
In other terms, the costs associated with sensing and adjusting metabolism can lead to the existence of an optimal value of $\beta$, i.e. of an optimal level of trade-off between $\avg{\mu}$ and $I$, such that (expectedly) higher values of ${\cal F}$ are possible only if the cost of encoding information about the environment in metabolic strategies gets smaller. (A similar idea was discussed in a different context in \cite{statmech}.)

\begin{figure}
\begin{center}
\includegraphics[width=\textwidth]{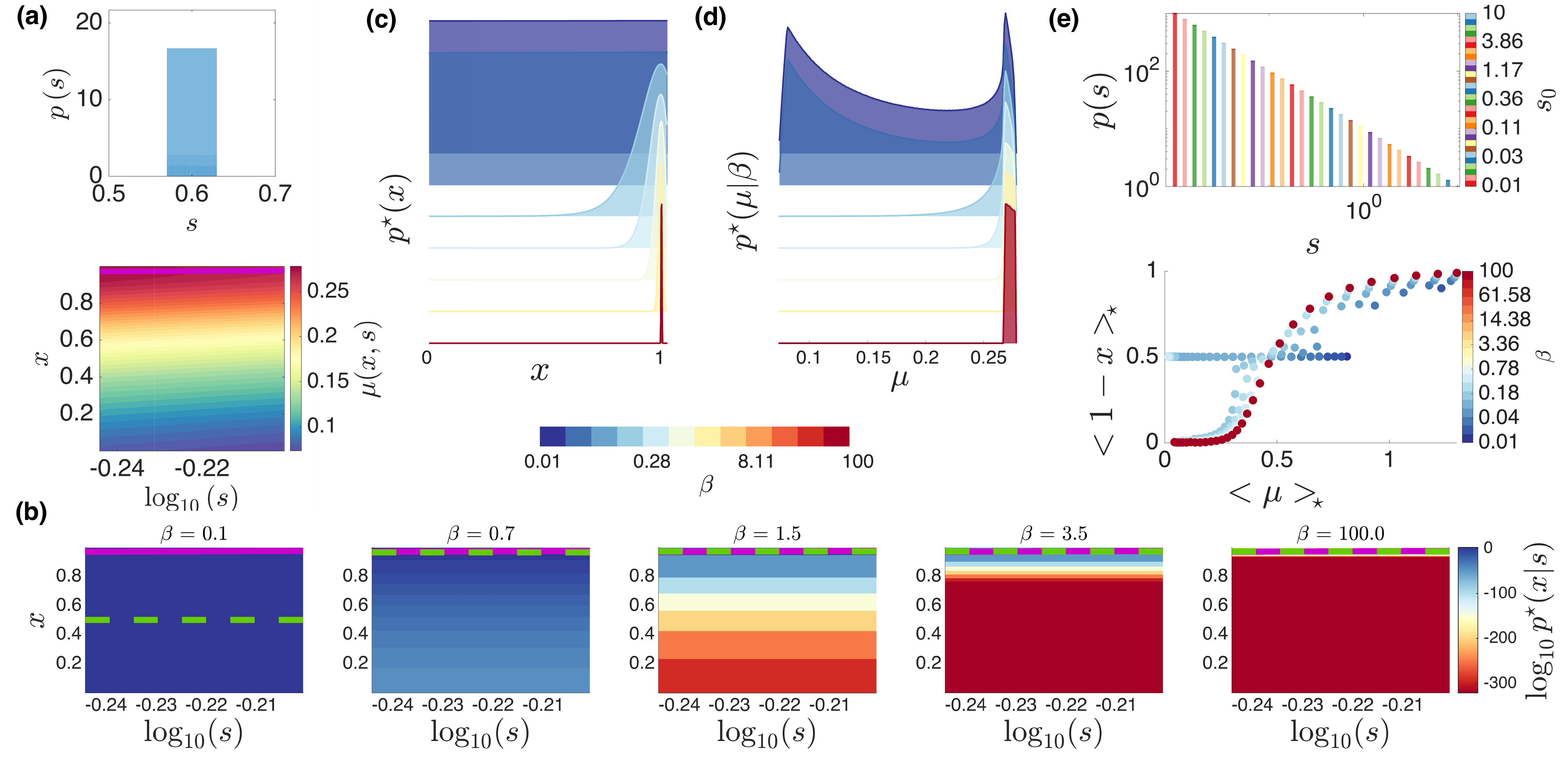}
\caption{(a) Probability density of the stress level $s$ in a controlled environment with $s_{0} = 0.6>s_c$ (top) and corresponding fitness landscape (bottom). Magenta line: growth-maximizing value of $x$ for each $s$. (b) Optimal choice for $p^{\star}\left(x|s\right)$ for different values of $\beta$. Magenta line: growth-maximizing value of $x$ for each $s$. Green line: mean value of $x$ for the optimal growth strategy as a function of $s$. (c) Marginal probability density $p^{\star}\left(x\right)$ for different values of $\beta$. (d) Growth rate distribution at optimality, $p^{\star}\left(\mu | \beta \right)$, for different values of $\beta$. (e, top) Several stress distributions centered at different values of $s_{0}$. (e, bottom) Behavior of the quantity $\avg{1-x}_\star$ as a function of the fitness for several values of $\beta$ and $s_0$ (for fixed $\beta$, the fitness increases as $s_0$ decreases). A value of $\avg{1-x}_\star\simeq 0$ indicates a strong preference for $R$ (respiration); a value of $\avg{1-x}_\star\simeq 1$ points instead towards massive use of $F$ (fermentation). \label{fig:fig2}}
\end{center}
\end{figure}

\subsection{Tightly controlled stress levels}

We start from the simple case where the stress level $s$ is tightly regulated, as in a lab setting where the nutrient level is externally controlled with high precision. Specifically, we assume that $p(s)$ is uniform and centered at a value $s_{0}$, and that $s$ can only vary by 5\% with equal probability around $s_0$ (Figure \ref{fig:fig2}a, top panel). The ensuing landscape of $\mu(x,s)$ is shown in Figure \ref{fig:fig2}a, bottom panel. One sees that, with our choice of $s_0$ ($s_0>s_c$, see (\ref{eq:x_hat})), growth is maximized for $x\simeq 1$ ($R$ strategy) for all values of $s$ (magenta line in Fig. \ref{fig:fig2}a). Figure \ref{fig:fig2}b reports the optimal strategy $p^\star(x|s)$ computed numerically for five different values of $\beta$. As expected, the distribution concentrates around the growth-maximizing value of $x$ for large enough $\beta$, while it becomes more and more uniform over the $[0,1]$ interval (with the mean $x$, green dashed line, getting closer and closer to $1/2$) as $\beta$ approaches zero. This gradual shift is also clearly visible in the behaviour of $p^\star(x)$, representing the distribution of metabolic strategies at optimality (Figure \ref{fig:fig2}c). Correspondingly, the growth-rate distribution $p^\star(\mu|\beta)$, which can be obtained numerically as
\begin{eqnarray}
p^\star(\mu|\beta)&=&\int ds\,p(s)\int dx\, p^\star(x|s)\,\delta[\mu(x,s)-\mu]~~,
\end{eqnarray}
is bimodal for small $\beta$ but then tends to get more and more concentrated around the maximum achievable values of $\mu(x,s)$ as $\beta$ increases (Figure \ref{fig:fig2}d). The residual variability seen at large $\beta$ reflects the fact that the stress level can only be known up to a finite precision (see Figure \ref{fig:fig2}a, top panel).

It is instructive to visualize together the results of a series of box-like environments $p(s)$ with different values of $s_0$, from very low to very high (see Figure \ref{fig:fig2}e, top panel), a setup that mimics an ensemble of experiments performed at different, exogenously controlled nutrient levels. As soon as $\beta$ is large enough, the mean response of the system (i.e. the mean value of $x$) changes from $x\simeq 0$ ($F$ strategy) in lower-stress media (faster growth) to $x\simeq 1$ ($R$ strategy) in higher-stress media (slower growth), see Figure \ref{fig:fig2}e, bottom panel. This scenario recapitulates qualitatively the well-known respiration-fermentation crossover marked by the onset of acetate excretion that is observed in experimental {\it E. coli} populations \cite{overflow}. Noticeably, however, it appears already for values $\beta$ well below the reference scale given by $1/\mu_{\max}$ (with $\mu_{\max}$ the fastest growth rate achievable in this medium,  approximately equal to 1/4 in this case). Only for values of $\beta$ much smaller than this does the crossover not develop. In other terms, in these simple environments, the crossover appears to be a feature of the optimal stochastic response $p^\star(x|s)$ that is extremely robust over $\beta$.

\begin{figure}[t!]
\begin{center}
\includegraphics[width=\textwidth]{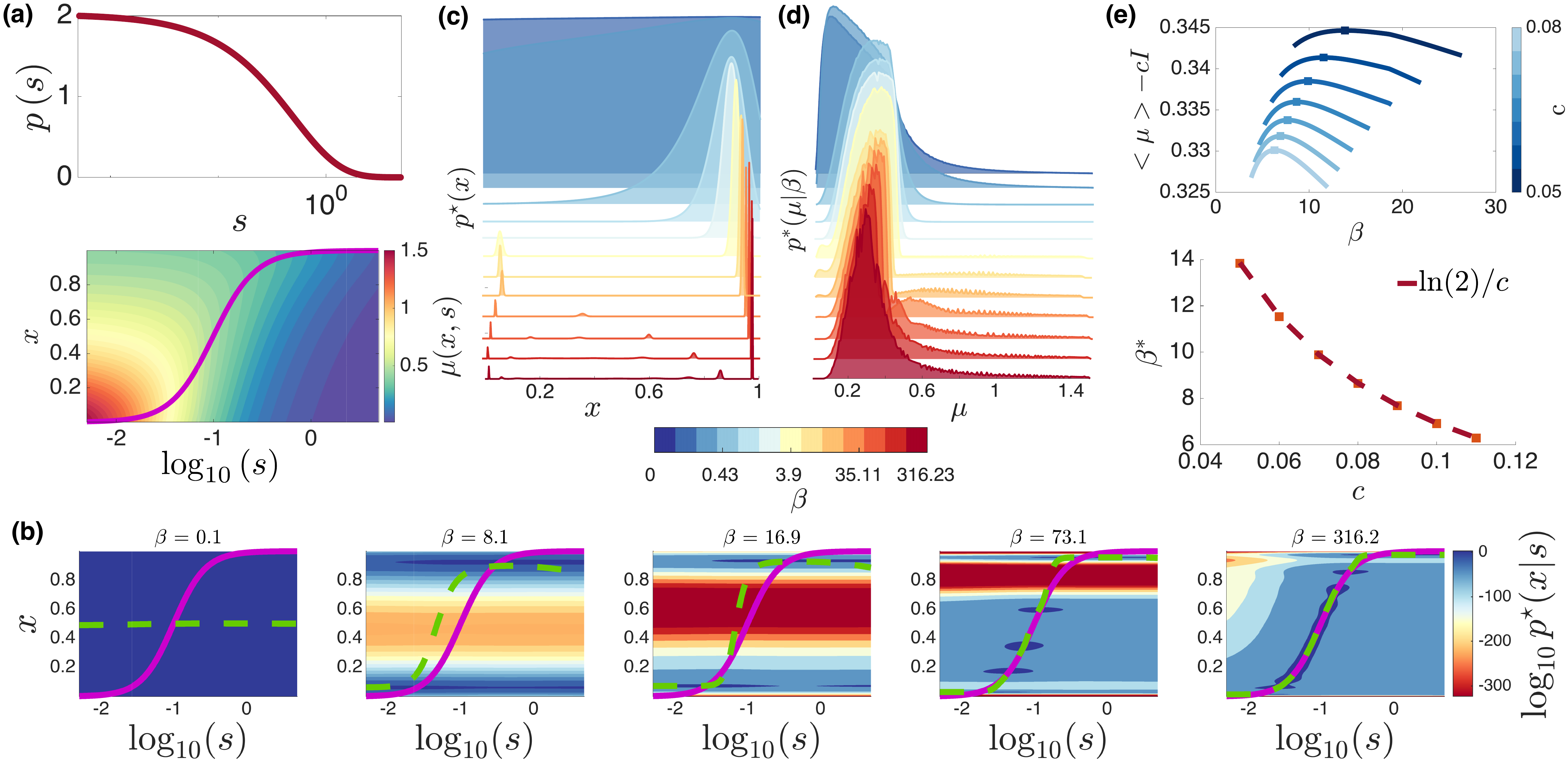}
\caption{(a) Probability density of the stress level $s$ in the `exponential environment' (top) $p\left(s\right) \propto e^{-s/s_{0}}$ for $s_{0} = 0.5$, $s \in \left[ 5 \cdot 10^{-3}, 5 \cdot 10^{0} \right]$ and corresponding fitness landscape (bottom). Magenta line: growth-maximizing value of $x$ for each $s$. (b) Optimal choice for $p^{\star}\left(x|s\right)$ for different values of $\beta$. Magenta line: growth-maximizing value of $x$ as a function of $s$. Green line: mean value of $x$ for the optimal choice rule at different values of $s$. (c) Marginal probability density $p^{\star}\left(x\right)$ for different values of $\beta$. (d) Growth rate distribution at optimality, $p^{\star}\left(\mu | \beta \right)$, for different values of $\beta$. (e, top) Overall fitness $\mathcal{F}$ (Eq. (\ref{resfit})) as a function of $\beta$ for several values of the information cost-per-bit $c$. (e, bottom) Optimal value of $\beta$ ($\beta^\star$) versus $c$: numerical results (markers) vs theoretical prediction (Eq. (\ref{betasta}), dashed line). \label{fig:fig4}}
\end{center}
\end{figure} 

\subsection{Exponentially distributed stress levels}

We now turn to the case in which the stress level is exponentially distributed. Specifically, we assume that $p(s)\propto \exp(-s/s_0)$ on the interval $\left[s_{\rm{min}},\, s_{\rm{max}} \right]$ (Figure \ref{fig:fig4}a, top panel). The corresponding fitness landscape $\mu(x,s)$ is shown in Figure \ref{fig:fig4}a, bottom panel, along with the growth-maximizing curve $x=\widehat{x}(s)$ (magenta line). One sees that $\widehat{x}(s)$ now shifts continuously from 0 ($F$ strategy) to 1 ($R$ strategy) as the stress level increases. Figure \ref{fig:fig3}b shows instead the optimal growth strategy $p^\star(x|s)$ for different values of $\beta$, with the green line representing the mean value of $x$ at each $s$. As expected, for small enough $\beta$, $p^\star(x|s)$ is roughly uniform over $[0,1]$. Upon increasing $\beta$, however, the distribution acquires a structure that more and more closely concentrates around $\widehat{x}(s)$. The qualitatively correct crossover is however already observed at values of $\beta$ of the order of $1/\mu_{\max}$. In turn, the distribution of $x$ at optimality, $p^\star(x)$, acquires a very strong bimodal character as $\beta$ increases (Figure \ref{fig:fig4}c). This clearly indicates that the coexistence of two sub-populations of fermenters ($x\simeq 0$) and respirators ($x\simeq 1$) provides the optimal metabolic response even in presence of limited resources to encode information about the environment into metabolic strategies, giving a quantitative representation to {\it bet-hedging} in the present model. Correspondingly, the growth-rate distribution (Figure \ref{fig:fig4}d) consistently displays large variability with a surprisingly weak dependence on $\beta$. Note that no bimodality is observed in $p^\star(\mu|\beta)$ despite the bimodality of $p^\star(x)$. 

It is important to remark that large fluctuations persist even in the $\beta\to\infty$ limit, when the information about the stress levels encoded in the metabolic strategy is maximal. In other terms, the globally optimal growth strategy ($\beta\to\infty$) in an exponentially-distributed medium requires a large degree of heterogeneity in growth rates. To better visualize the robustness of these outcomes with respect to $\beta$, we display in Figure \ref{fig:fig4}e the re-scaled fitness ${\cal F}$ defined in (\ref{resfit}) as a function of the `inverse temperature'. As anticipated, ${\cal F}$ has a maximum at intermediate value of $\beta$, related to the cost-per-bit $c$ of encoding information as shown in (\ref{betasta}) (bottom panel): higher costs imply lower optimal values of $\beta$. Noticeably, though, with the realistic parameters we used, the maximum of the fitness function is extremely flat. This suggests that, at least in principle, metabolic networks could be capable of generating optimal growth profiles that are robust to the amount of information about the environment encoded in the growth strategy. In this respect, questions concerning the fine-tuning of the value of $\beta$ might be less relevant for this problem than they likely are for the MaxEnt-related growth-heterogeneity trade-off discussed in \cite{growthvs,relationship,statmech}: relatively limited environmental cues could suffice to bias the distribution of metabolic strategies enough to generate a (stochastic) response close to the globally optimal one, i.e. that achievable by perfectly matching $x$ to the stress levels.

\subsection{Power-law distributed stress levels}

\begin{figure}[t!]
\begin{center}
\includegraphics[width=\textwidth]{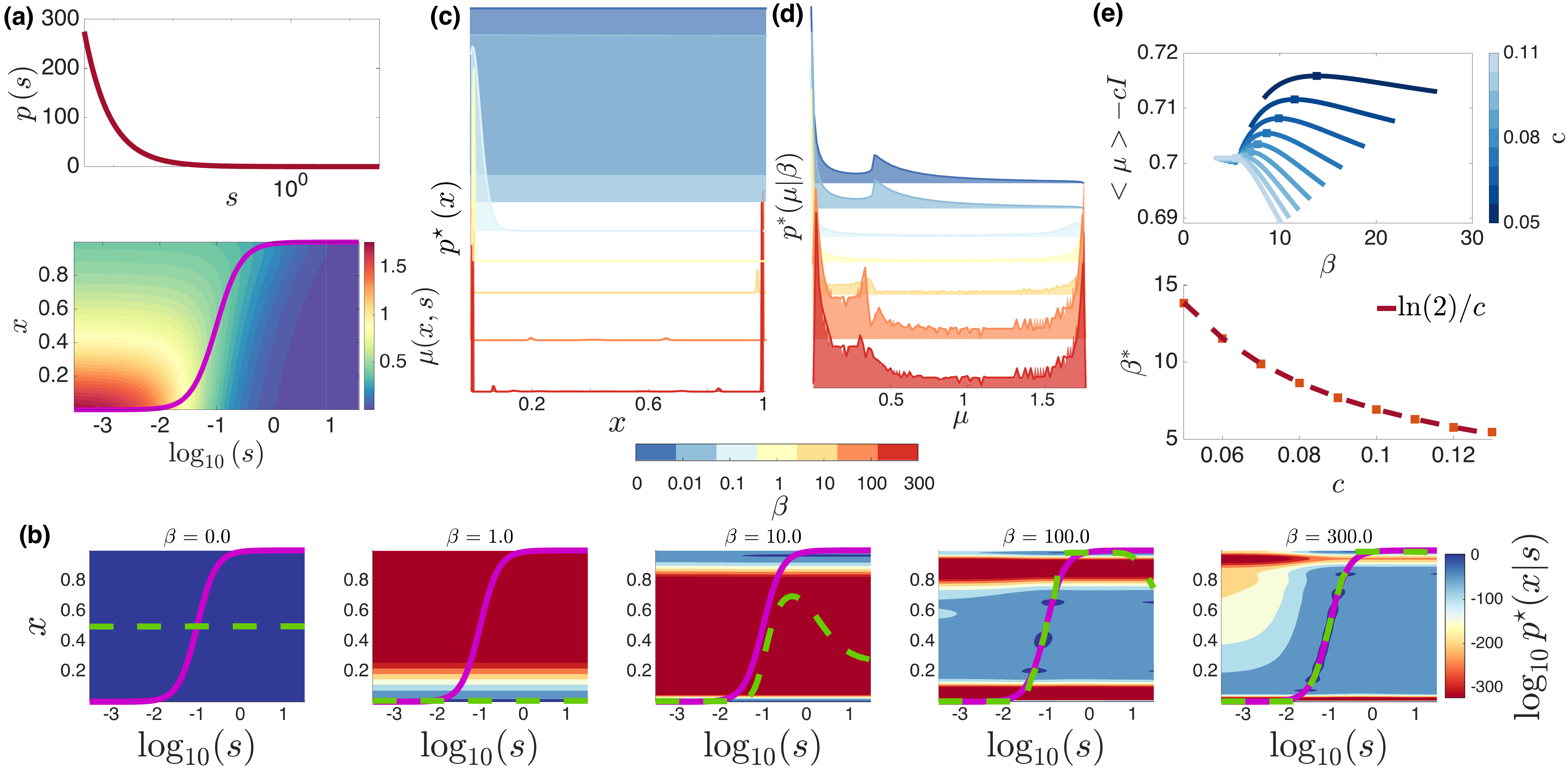}
\caption{(a) Probability density of the stress level $s$ in the `power-law environment' (top) and corresponding fitness landscape (bottom). Here, $s_{\min}= 3 \cdot 10^{-4}$ and $s_{\max} = 3 \cdot 10^{1}$. Magenta line: growth-maximizing value of $x$ for each $s$. (b) Optimal choice for $p^{\star}\left(x|s\right)$ for different values of $\beta$. Magenta line: growth-maximizing value of $x$ as a function of $s$. Green line: mean value of $x$ for the optimal choice rule at different values of $s$. (c) Marginal probability density $p^{\star}\left(x\right)$ for different values of $\beta$. (d) Growth rate distribution at optimality, $p^{\star}\left(\mu | \beta \right)$, for different values of $\beta$. (e, top) Overall metabolic objective function $\mathcal{F}$ (Eq. (\ref{resfit})) as a function of $\beta$ for several value of the information cost-per-bit $c$. (e, bottom) Optimal value of $\beta$ ($\beta^\star$) versus $c$: numerical results (markers) vs theoretical prediction (Eq. (\ref{betasta}), dashed line). \label{fig:fig5}}
\end{center}
\end{figure}

\begin{figure}
\begin{center}
\includegraphics[width=\textwidth]{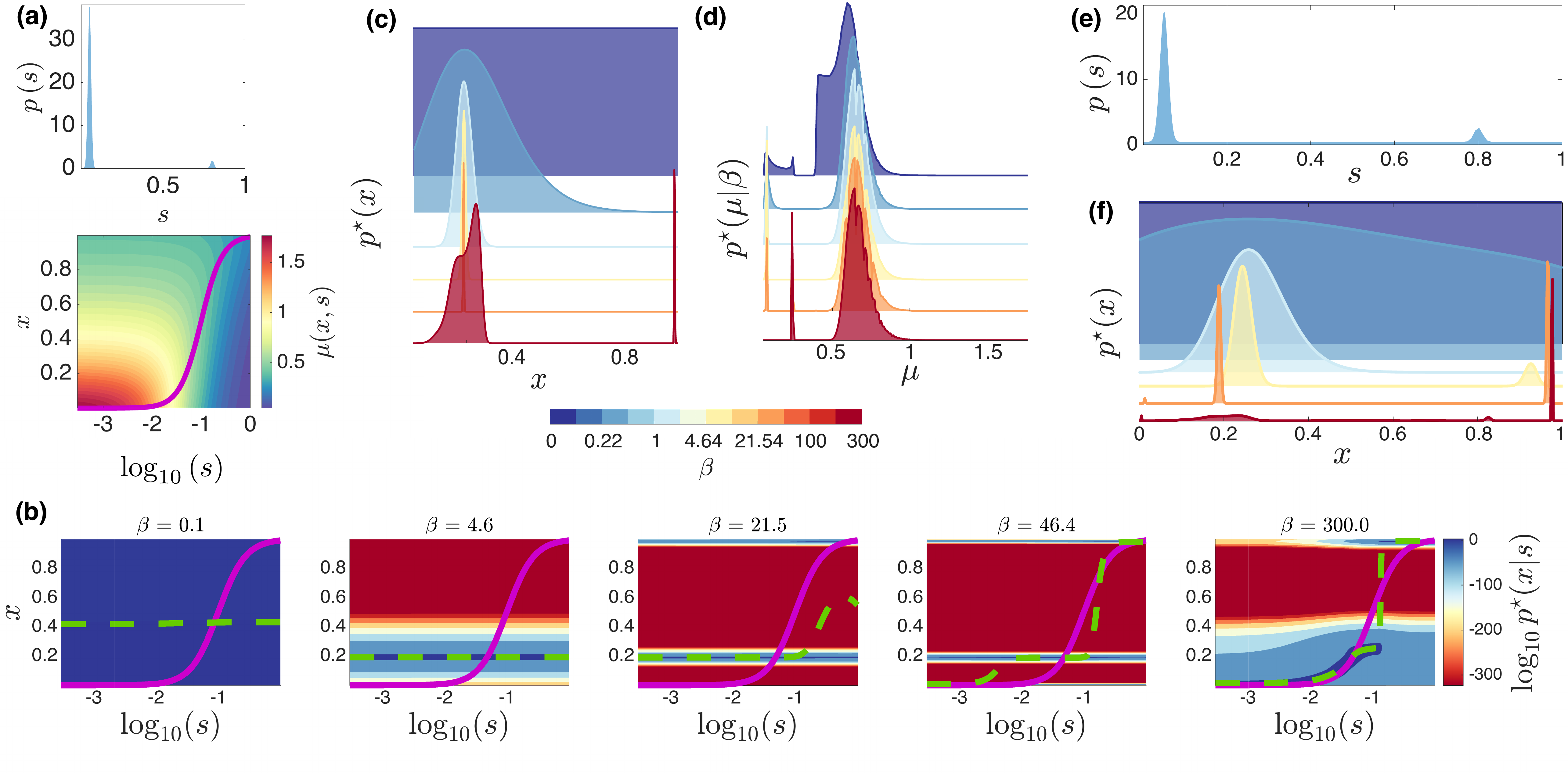}
\caption{(a) Probability density of the stress level $s$ in a two-state environment (top) and corresponding fitness landscape (bottom). Magenta line: growth-maximizing value of $x$ for each $s$. (b) Optimal choice for $p^{\star}\left(x|s\right)$ for different values of $\beta$. Magenta line: growth-maximizing value of $x$ for each $s$. Green line: mean value of $x$ for the optimal choice rule at different as a function of $s$. (c) Marginal probability density $p^{\star}\left(x\right)$ for different values of $\beta$. (d) Growth rate distribution at optimality, $p^{\star}\left(\mu | \beta \right)$, for different values of $\beta$. (e) Alternative bimodal stress distribution with an extended background of stress levels with non-zero probability. (f) Corresponding growth rate distribution at optimality, $p^{\star}\left(\mu | \beta \right)$, for different values of $\beta$.\label{fig:fig3}}
\end{center}
\end{figure}

A yet more complex scenario is that in which the distribution of stress levels has a power-law behaviour of the type $p\left( s\right) \propto 1/s$ on the interval $s \in \left[s_{\rm{min}}, s_{\rm{max}} \right]$ (Figure \ref{fig:fig5}a, top panel, with the corresponding growth rate landscape $\mu(x,s)$ reported in the bottom panel). Results for this case are shown in Figures \ref{fig:fig5}b-e. In brief, the optimal growth strategy $p^\star(x|s)$, which is uniform for sufficiently small $\beta$, matches the globally optimal one (i.e. $\widehat{x}(s)$, represented by the magenta line in Figure \ref{fig:fig5}a) more and more closely as $\beta$ increases (Figure \ref{fig:fig5}b). The distribution of $x$ at optimality (Figure \ref{fig:fig5}c) again underscores a non-trivial structuring of the population, which separates into well-defined groups of faster-growing fermenters ($x\simeq 0$) and slower-growing respiratos ($x\simeq 1$). Remarkably, the separation already occurs effectively at values of $\beta$ comparable with $1/\mu_{\max}$, with $\mu_{\max}$ the fastest growth rate achievable in the medium. At odds with the exponential case, however, the growth-rate distribution now displays bimodality robustly over $\beta$ (see Figure \ref{fig:fig5}d). Finally, as in the exponential case, the value of $\beta$ yielding the maximum fitness $\mathcal{F}$ decreases with the cost-per-bit $c$, and the maximum again appears to get flatter as $c$ decreases (Figure \ref{fig:fig5}e), suggesting that limited resources may suffice to fine-tune $\beta$ so as to achieve responses that are effectively optimal.

\subsection{Two-state environments}

Figure \ref{fig:fig3} displays results obtained for the case of a bimodal environment in which high-probability favourable conditions coexist with low-probability very adverse ones (Figure \ref{fig:fig3}a), as e.g. in experiments probing bacterial resistance to stress via antibiotic cycles \cite{theimportance}. Here, 95 \% of the distribution is concentrated around $s_{0} = 0.05<s_c$, whereas high stress levels take values around $s_{1} = 0.8>s_c$\footnote{The low-stress (high-stress) region is obtained by truncating in the interval $\left[0,\,0.1\right]$ ($\left[0.7,\,0.9\right]$) a normal density of mean $0.05$ ($0.8$) and standard deviation $10^{-2}$ ($10^{-2}$). For $s \in \left(0.1,\,0.7 \right)$, $p\left(s\right) = 0$. In other terms, intermediate values of $s$ are strictly prohibited.}. According to the growth-rate landscape (Figure \ref{fig:fig3}a, bottom panel), while the low-cost strategy ($x=0$) is favored in rich conditions, high-yield metabolism ($x=1$) provides the optimal response under stress (magenta line in Figure \ref{fig:fig3}a, bottom panel). Optimal adaptation to a strict bimodal environment therefore requires the ability to neatly separate the two phenotypes. Indeed, as $\beta$ increases, the optimal strategy selection rule $p^\star(x|s)$ reproduces more and more closely the optimal response (Figure \ref{fig:fig3}b), while the population gradually acquires a structure in terms of $x$. As in previous cases, as soon as $\beta$ is sufficiently large one observes the formation of two sub-populations (Figure \ref{fig:fig3}c): one formed by cells that use a mixed strategy with a stronger fermentative component (smaller $x$), the other formed by a small but robust group of strict respirators ($x\simeq 1$). This partitioning is reflected in turn in the distribution of growth rates (Figure \ref{fig:fig3}d), which displays a marked bimodal character even for large $\beta$, with lower growth rates associated with respirators.  (It should however be noted that, in this environment, the background growth-rate landscape is effectively bimodal, as seen from the $\beta\to 0$ limit in Figure \ref{fig:fig3}d.) These results paint a direct metabolic realization of the {\it bet-hedging} scenario, with respirators forming the group of `persisters' \cite{theimportance}. 

Noticeably, the dependence on $\beta$ in this environment appears to be stronger than in previous cases, suggesting that optimal adaptation to a two-state medium requires the encoding of a considerable amount of information about stress levels in the metabolic strategy. In view of this, it is reasonable to ask how the presence of a broad, low-probability background of stress levels along with the two `peaks' at low and high stress would change optimal adaptation. Remarkably, results shown in Figure \ref{fig:fig3}e suggest that this scenario mainly affects the faster-growing sub-population\footnote{To obtain the probability density shown in Figure \ref{fig:fig3}e (top panel), we first removed the truncations used for  Figure \ref{fig:fig3}a, then added a uniform distribution in the interval $\left[0,\,1.0\right]$ multiplied by $0.8$, and finally normalized the overall $p\left(s\right)$.}. Indeed, as $\beta$ grows, the optimal population tends to exploit intermediate values of $x$ in a more and more refined way by depleting the sub-population at smaller values of $x$ (Figure \ref{fig:fig3}f, to be compared with Figure \ref{fig:fig3}c). In other terms, the presence of a robust sub-population of slower-growing respirators is crucial for optimal adaptation.

\begin{figure}
\begin{center}
\includegraphics[width=\textwidth]{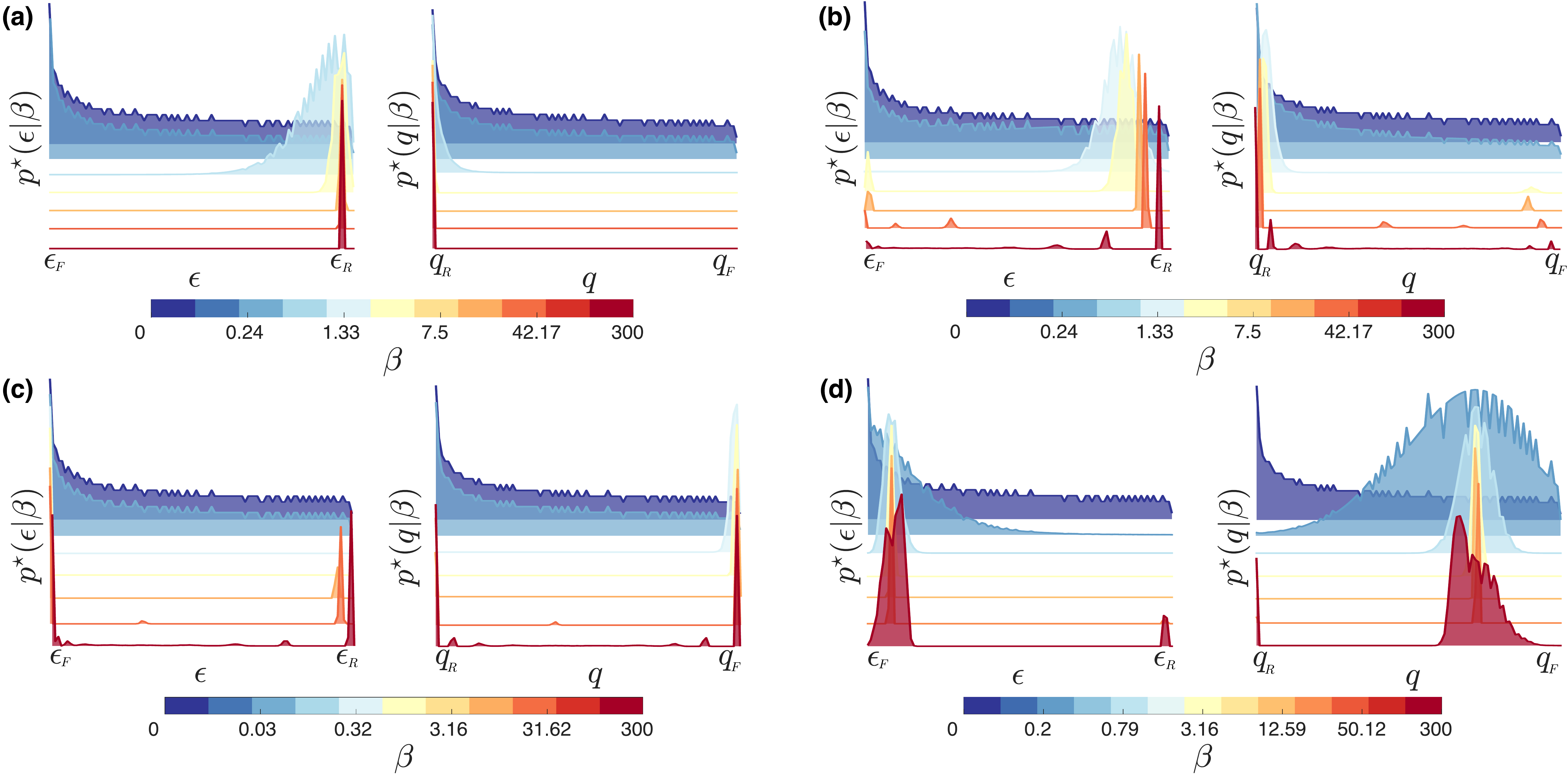}
\caption{Numerical results for the probability densities of the specific proteome cost $\eps$ (left plot in each panel) and the specific uptake $q$ (right plot in each panel) for different values of $\beta$. Panels (a)-(d) correspond respectively to the cases of a tightly controlled (a), exponential (b), power-law (c) or strict bimodal (d) stress distribution discussed in Figures \ref{fig:fig2}-\ref{fig:fig3}. The lower and upper bounds for $\eps$ and $q$ are fixed to $\eps_{F} = 0.1$ and $q_{R} = 1$, and to $\eps_{R} = 1$ and $q_{F} = 10$, respectively. \label{fig:fig6}}
\end{center}
\end{figure}

\subsection{A closer look at optimal growth strategies}

It is finally instructive to revisit the optimal phenotypic distributions found in the four types of environments discussed above in terms of the biologically relevant variables, namely the specific intake $q$ and the specific proteome cost $\eps$. The distribution of these quantities can be obtained numerically from the optimal growth strategy $p^\star(x|s)$ and the definitions (\ref{eq:q_q}) and (\ref{eq:q_eps}) as
\begin{eqnarray}
p^\star(q |\beta)&=&\int ds\,p(s)\int dx\, p^\star(x|s)\,\delta[q(x)-q]~~, \\
p^\star(\eps |\beta)&=&\int ds\,p(s)\int dx\, p^\star(x|s)\,\delta[\eps(x)-\eps]~~. \\
\end{eqnarray} 
Results are reported in Figure \ref{fig:fig6}a-d. One clearly sees how, expectedly, optimal adaptation to a tightly controlled medium gradually concentrates the metabolic phenotypes around the growth-maximizing one as $\beta$ increases (Figure \ref{fig:fig6}a). In all other (more complex) cases, however, the distributions of metabolic phenotypes effectively define clear sub-populations already for values of $\beta$ comparable with $1/\mu_{\max}$. Based on the previous results, this process by itself can lead to an outcome close to the globally optimum one ($\beta\to\infty$) in an exponential environment (Figure \ref{fig:fig6}b) as well as in a power-law environment (Figure \ref{fig:fig6}c). Indeed in both cases phenotypic distributions do not change substantially as $\beta$ is further increased. The bimodal environment again appears to be more sensitive to the value of $\beta$ (i.e. to the amounts of resources available), as its corresponding phenotypic distribution undergoes considerable refinements as $\beta$ gets larger and larger (Figure \ref{fig:fig6}d). In this respect, it is fair to say that bimodal media pose an especially complex adaptation problem for cells that strive to optimize growth.

\section{Discussion}

The standard theoretical framework to describe exponentially growing microbial populations mostly relies on the assumption that cells aim at maximizing their growth rate in any given medium. This idea should however be reconciled with the empirical fact that bacterial populations display remarkable cell-to-cell diversity even in terms of single-cell growth rates. Such a question is especially pressing in the context of metabolic networks, where optimality normally implies a maximal reduction of variability (e.g. as in Flux-Balance-Analysis and related models). Here we attempt to bridge optimality and variability in metabolism through the fact that metabolic strategies need to encode information about the growth medium. More precisely, optimal strategies for microbial growth in complex (random) media depend on the amount of information about the environment that can be encoded in the key metabolic variable(s). By framing this idea within a simple information-theory perspective, we have studied the optimal growth strategies for a coarse-grained but experiment-inspired model of growth metabolism subject to random stress levels. The key parameter $\beta\geq 0$ modulates, in essence, how precisely metabolic variables can be tuned in response to the stress levels or, equivalently for our purposes, the amount of cellular resources available to tune the metabolic response to the environmental conditions. By changing $\beta$, one passes from optimal strategies that are completely insensitive to the composition of the medium ($\beta=0$) to strategies that are perfectly matched to the structure of the environment ($\beta\to\infty$). We have then analyzed the scenarios generated by these strategies in different types of environments. The general lesson is that, when cellular resources are limited, metabolic strategies for optimal growth require some degree of heterogeneity even in tightly controlled media. In addition, we find that, in realistic scenarios, outcomes close to those achievable with unlimited resources are often attained effectively with a modest amount of fine-tuning or, equivalently, with limited information about the environment. In other terms, optimal metabolic strategies for growth may be effectively robust with respect to the amounts of cellular resources available.

From a purely theoretical perspective, models addressing the adaptation of populations to unpredictable external conditions have been studied for several decades. In our view, despite lacking an explicit dynamics, our approach is conceptually related to that of  \cite{theevo}, where heterogeneities in the phenotype distribution emerge as evolutionarily stable strategies to cope with fluctuating environments. Likewise, the `exploration-exploitation' paradigm discussed in \cite{exex} leads to a similar outcome in terms of growth rate distributions, albeit in a highly stylized framework. A dynamical counterpart of the static scenario discussed here indeed shows that optimal adaptation can be defined even for dynamically varying stress levels (forthcoming). 
On the other hand, heterogeneity of growth rates (and indirectly of metabolic phenotypes) has been repeatedly observed in experiments probing actual cell populations subject to different stressors. Large variability in growth-rate distributions characterizes for instance microbial communities in fluctuating conditions \cite{theimportance,microbialbet,diversityand}, where heterogeneity induced e.g. by changes in environmental cues coexists with medium-independent phenotypic variability due to stochasticity in gene expression. By expanding the pool of genotypes present within a population, the latter plays a key role for its long-term adaptation \cite{afunctional}. Obtaining a direct quantitative picture of the the metabolic strategies of individual cells in a live population is instead experimentally challenging. Recent work integrating high-resolution techniques for local microenvironment sensing with statistical inference has however allowed to quantitatively assess the presence of different metabolic phenotypes in a population of cells that engineers its own environment \cite{probingsingle}. Progress along these lines will hopefully shed more light on the distributions of biologically relevant variables in phenotypically heterogeneous populations.

The results we presented here suggest that, in any medium, fitness is maximized by an optimal distribution of phenotypes. To understand how close bacterial populations come to these optimal strategies, one needs to port our modeling framework to more realistic models of metabolic networks. It would be especially interesting, in our view, to compare the predictions based on optimal growth driven by information encoded about the environment with those obtained within the fitness-heterogeneity trade-off scenario of \cite{growthvs,relationship,statmech}. These approaches rely on the MaxEnt principle to characterize optimal phenotypic distributions in terms of a fitness-variability trade-off, and in turn provide a route to inferring how close real microbial populations are to optimality. So far, this issue has only been explored for populations living in tightly-controlled media \cite{relationship}. Naturally, however, things may be environment-dependent. With proper data, and by integrating the current framework with the MaxEnt models introduced in \cite{growthvs}, one could potentially establish how much of the observed variability is due to adaptation to complex environments versus stochastic bet-hedging.

A few other generalizations and open issues could be of further interest. First, we assumed the value of $\nu$ (Equation (\ref{eq:q_eps})) to be given, representing a property of the metabolic and regulatory networks underlying the $q-\epsilon$ trade-off. Clearly, results depend on the choice of $\nu$, albeit weakly if $\nu$ varies in a realistic range. Important and potentially insightful question however relate to (i) whether the value of $\nu$ can, in some sense, be optimized upon, and to (ii) how close empirical values of $\nu$ are to such optima. An information-theoretic framework similar to that employed here might be suited to tackle this issue. Second, in our formulation we have assumed that bacteria have knowledge of the ensemble of conditions in which they grow, i.e. of $p(s)$. This model can however be generalized to the case in which bacteria have to learn the ensemble over time. From an information theory perspective, a dynamical approach is likely to yield yet more insight into the idea of `optimal bacterial growth'. Finally, it may be noted that our minimal model of growth completely disregards maintenance costs, namely the fact that cells can only grow when certain basal (and possibly growth-rate dependent) metabolic requirements are satisfied. 
While results presented here can be seen as the low-maintenance limit of this more general case, optimal strategies and growth rate distributions are bound to be affected by non-trivial maintenance costs, especially in heterogeneous media. A detailed study of these aspects would clarify the interplay between optimal growth, maintenance and stress levels in complex environments.

\section*{Data availability statement}
The data that support the findings of this study are openly available at the following URL \url{https://github.com/anna-pa-m/OptMetStrategy}.

\section*{References}


\appendix

\section{Derivation of the optimal distribution}

Our goal is to determine the conditional distribution $p\left(x|s\right)$ that maximizes the expectation value of the growth rate --computed with respect to the joint probability density of the degree of freedom $x$ and of the stress $s$, i.e. $p\left(x,s\right)$-- when the mutual information regarding the stress encoded in the conditional distribution is limited to a constant; that is, we want to solve
\begin{eqnarray}
\max_{p(x|s)}\avg{\mu}~~~\textrm{subject to}~~~ 
I=\mathrm{constant}~~.
\end{eqnarray}
To this end, we introduce, within the formalism of Lagrange multipliers, the Lagrangian function 
\begin{eqnarray*}
\mathcal{L}\left[p\left(x|s\right)\right] & = & \int ds\,p\left(s\right)\int dx\,p\left(x|s\right)\mu\left(x,s\right)\\
 &  & +\gamma\left[\int ds\,p\left(s\right)\int dx\,p\left(x|s\right)\left[\log_{2}p\left(x|s\right)-\log_{2}p\left(x\right)\right]-I\right]\\
 &  & +\int ds\,\left[\lambda\left(s\right)\int dx\,p\left(x|s\right)-1\right]~~,
\end{eqnarray*}
where $\gamma$ is the Lagrange multiplier associated with the constraint on the mutual information between $s$ and $x$, and $\lambda\left(s\right)$ ensures that for any value of $s$, the conditional distribution $p\left(x|s\right)$ is normalized to $1$. Computing the functional derivative one gets
\begin{eqnarray*}
\frac{\delta\mathcal{L}}{\delta p\left(x|s\right)} & = & \int ds'\int dx'\delta\left(x-x'\right)\delta\left(s-s'\right)p\left(s'\right)\mu\left(x',s'\right)\\
 &  & +\gamma\left[\int ds'\int dx'p\left(s'\right)\log_{2}\frac{p\left(x'|s'\right)}{p\left(x'\right)}\delta\left(x-x'\right)\delta\left(s-s'\right)\right]\\
 &  & +\gamma\left[\int ds'\int dx'p\left(s'\right)p\left(x'|s'\right)\frac{1}{p\left(x'|s'\right)}\delta\left(x-x'\right)\delta\left(s-s'\right)\right]\\
 &  & -\gamma\left[\int ds'\int dx'p\left(s'\right)p\left(x'|s'\right)\frac{1}{p\left(x'\right)}\frac{\delta p\left(x'\right)}{\delta p\left(x|s\right)}\right]\\
 &  & +\int ds'\,dx'\,\lambda\left(s'\right)\delta\left(s'-s\right)\delta\left(x-x'\right)~~.
\end{eqnarray*}
The remaining integrations involving Dirac $\delta$-functions are easily performed, yielding
\begin{eqnarray}
\frac{\delta\mathcal{L}}{\delta p\left(x|s\right)} & = & p\left(s\right)\mu\left(x,s\right)+\gamma\left[p\left(s\right)\log_{2}\frac{p\left(x|s\right)}{p\left(x\right)}+p\left(s\right)-I_{0}\right]+\lambda\left(s\right)~~,
\end{eqnarray}
where 
\begin{eqnarray}
I_{0}=\int ds'\int dx'p\left(s'\right)p\left(x'|s'\right)\frac{1}{p\left(x'\right)}\frac{\delta p\left(x'\right)}{\delta p\left(x|s\right)}~~.
\end{eqnarray}
Note however that $p\left(x\right) = \int ds\,p\left(x,s\right)=\int ds\,p\left(x|s\right)p\left(s\right)$. Therefore
\begin{eqnarray*}
I_{0} & = & \int ds'\,\int dx'\,p\left(s'\right)p\left(x'|s'\right)\frac{1}{p\left(x'\right)}\int ds''p\left(s''\right)\delta\left(s''-s\right)\delta\left(x'-x\right)\\
 & = & p\left(s\right)\int dx'\frac{1}{p\left(x'\right)}\int ds'\,p\left(s'\right)p\left(x'|s'\right)\delta\left(x-x'\right)\\
 & = & p\left(s\right)
\end{eqnarray*}
Setting the derivative of the Lagrangian to zero leads to the condition
\begin{eqnarray}
p\left(s\right)\left[\mu\left(x,s\right)+\gamma\log_{2}\frac{p\left(x|s\right)}{p\left(x\right)}+\gamma-\gamma\right]+\lambda\left(s\right)=0~~,
\end{eqnarray}
from which one immediately gets
\begin{eqnarray}\label{a5}
p^{\star}\left(x|s\right) & \propto & p^{\star}\left(x\right)e^{\beta\mu\left(x,s\right)+\beta\frac{\lambda\left(s\right)}{p\left(s\right)}}~~,
\end{eqnarray}
where here we have defined $\beta=-1/\gamma$. Upon imposing the normalization condition
\begin{eqnarray}
\int dx\,p^{\star}\left(x|s\right) & = & 1\qquad\forall s~~,
\end{eqnarray}
the value of $\lambda\left(s\right)$ can be straightforwardly determined to be
\begin{eqnarray}
\lambda\left(s\right) & = & \beta^{-1}p\left(s\right)\log\left[\int dx\,p^{\star}\left(x\right)e^{\beta\mu\left(x,s\right)}\right]^{-1}~~.
\end{eqnarray}
Substituting this into (\ref{a5}) one arrives at Equations (\ref{pags}) and (\ref{pags2}), where $N\left(s,\beta\right)=\int dx\,p^{\star}\left(x\right)e^{\beta\mu\left(x,s\right)}$.

\end{document}